\documentclass[conference, 10pt]{IEEEtran}

\ifCLASSINFOpdf
\else
\fi

\usepackage{graphicx}
\usepackage{amsmath}
\usepackage{color}
\usepackage{placeins}
\usepackage{float}
\usepackage{tabularx,colortbl}
\usepackage{psfrag}
\usepackage{epsfig}
\usepackage{subfigure}
\usepackage{multicol}
\usepackage{cite}
\usepackage[usenames,dvipsnames]{xcolor}
\usepackage{jabbrv}

\begin{document}
\title{Spacetime Processing Metasurfaces: \\ GSTC Synthesis \\ and Prospective Applications}

\author{\IEEEauthorblockN{Nima Chamanara, Yousef Vahabzadeh, Karim Achouri, and Christophe Caloz}
\IEEEauthorblockA{Polytechnique Montr\'{e}al,\\ Montr\'{e}al, Qu\'{e}bec H3T 1J4, Canada. \\e-mail: nima.chamanara@polymtl.ca}
}

\maketitle

\begin{abstract}
%\boldmath
The paper presents the general concept of spacetime processing metasurfaces, synthesized by generalized sheet transition conditions (GSTCs). It is shown that such metasurfaces can perform multiple simultaneous spatio-temporal processing transformations on incident electromagnetic waves. A time-reversal space-generalized-refraction metasurface and a multi-time-space-differentiating metasurfaces are presented as applications of the general spacetime processing metasurface concept.
\end{abstract}


\IEEEpeerreviewmaketitle

\section{Introduction}

Metasurfaces are periodic or aperiodic arrays of scattering elements arranged in a subwavelengthly thin substrate~\cite{holloway2012overview}. They transform electromagnetic waves in unusual ways and they have already been demonstrated in a diversity of applications, including polarization transformers and rotators~\cite{niemi2013synthesis,kodera2011artificial}, perfect absorbers~\cite{ra2013total}, beam transformers~\cite{salem2014manipulating}, electromagnetic interferometers and transistors~ \cite{achouri2015metasurface} and nonreciprocal metasurfaces~\cite{kodera2011artificial,Sounas_TAP_01_2013,hadad2015space,Khan_ISAP_11_2015}.

Designing metasurfaces is generally a challenging inverse problem. Synthesis techniques include heuristic approaches, and more general techniques such as momentum transformation~\cite{salem2014manipulating} and the general sheet transition condition (GSTC) synthesis approach~\cite{idemen1987boundary, achouri2014general}. The latter, that provides the general bianisotropic surface susceptibility tensors of the metasurface, is a powerful design technique: 1)~It is exact; 2)~It is general, transforming arbitrary incident waves into arbitrary reflected and transmitted waves, 3)~it often admits closed-form solutions, 4)~it provides deep insight into the physics of the transformations, 5)~it allows multiple (at least up to 4) simultaneous and independent transformations. However, this method has been so far restricted to monochromatic and space-only-varying wave transformations~\cite{achouri2014general}.

This paper extends the GSTC susceptibility technique to
space-and-time-varying metasurfaces and presents two corresponding applications. Moreover, it is noted that such a spacetime-varying metasurface also applies to the spatio-temporal processing of multiple simultaneous waves. 

The organization of the paper is as follows. The GSTC approach of spacetime-varying metasurfaces is described in Sec.~\ref{sec:gstc-metasurfs}. Two illustrative examples are presented in Sec.~\ref{sec:examples}. Conclusions are given in Sec.~\ref{sec:concl}.

\section{Time-Varying GSTC-based Metasurfaces} \label{sec:gstc-metasurfs}

Metasurfaces can be effectively modeled as a zero-thickness sheet that creates a discontinuity in the electromagnetic field. Such a discontinuity can be most generally described by generalized sheet transitions conditions (GSTCs)~\cite{idemen1987boundary}. In the time domain, GSTCs read

\begin{equation}\label{eq:GSTC_H}
\mathbf{n}\times\left(\mathbf{H^{+}}-\mathbf{H^{-}}\right)=\mathbf{J_{0}}+\frac{\partial}{\partial t}\mathbf{P}_{0\text{t}}-\frac{1}{\mu_{0}}\mathbf{n}\times\nabla M_{0\text{n}},
\end{equation}

\begin{equation}\label{eq:GSTC_E}
\mathbf{n}\times\left(\mathbf{E^{+}}-\mathbf{E^{-}}\right)=-\mathbf{K_{0}}-\frac{\partial}{\partial t}\mathbf{M}_{0\text{t}}-\frac{1}{\epsilon_{0}}\mathbf{n}\times\nabla P_{0\text{n}},
\end{equation}

\noindent
where the $+$ and $-$ superscript denote the fields specified on both sides of the metasurface ($-$: incident and reflected, $+$: transmitted), where the t and n subscripts denote the tangential and normal components, respectively, and where the $0$ subscript denote the first-order Dirac delta terms of the 2D discontinuity.

The conditions~\eqref{eq:GSTC_H} and~\eqref{eq:GSTC_E} are valid for any time-continuous electromagnetic waves. Therefore, one may specify arbitrary time-varying, in addition to spatially-discontinuous, electromagnetic fields on the two sides of the metasurface, $\mathbf{E^\pm}(\mathbf{r},t)$ and $\mathbf{H^\pm}(\mathbf{r},t)$. As a result, one finds the corresponding time-varying, in addition to space-varying, electric and magnetic polarization densities that satisfy these conditions~$\mathbf{P}_{0\text{t}}(\mathbf{r},t)$ and~$\mathbf{M}_{0\text{t}}(\mathbf{r},t)$. The time-varying and space-varying susceptibilities, $\overline{\overline{\chi}}_\text{uv}(\mathbf{r},t)$ where uv represents the 4 combinations ee, mm, em and me, can then be found through the averages of the electric and magnetic fields on the metasurface at each time instant, following a similar procedure as in~\cite{achouri2014general}.

\section{Illustrative Examples} \label{sec:examples}

A spacetime-varying metasurface can in principle arbitrarily transform the spatial and temporal frequency contents of any incident electromagnetic wave. We shall present here two illustrative application examples.

\begin{figure}[ht!]
\psfrag{m}[c][c][0.9]{ }
\psfrag{i}[l][c][0.9]{$\mathbf{E}^I(t)$}
\psfrag{t}[c][c][0.9]{$\mathbf{E}^T(t)$}
\includegraphics[width=0.9\columnwidth]{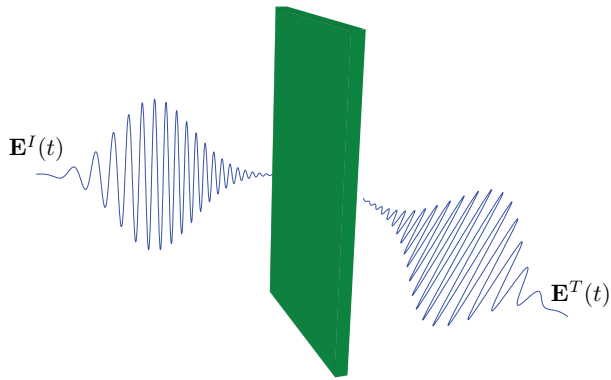}
\caption{Time-reversal generalized-refractive spacetime metasurface. A normally incident chirped pulse is time-reversed and refracted under an oblique angle.}
\label{fig:metatsurf_chirp_TR}
\end{figure}

The first application is depicted in Fig.~\ref{fig:metatsurf_chirp_TR}. A chirped Gaussian electromagnetic pulse is impinging on the spacetime-varying metasurface and the metasurface is specified to transform this wave into a reversed-chirped wave refracted in a different direction, and with zero reflection. The corresponding spacetime-varying metasurface susceptibility tensors are found following the procedure outlined in the previous section, leading exactly to the transformation shown in the figure. Note that the spatio-temporal frequency contents of the input chirp are simultaneously transformed by the metasurface: the normally incident electromagnetic pulse is time reversed (temporal frequency transformation) and transmitted at an oblique angle (spatial frequency transformation).

\begin{figure}[ht!]
\psfrag{m}[c][c][0.9]{ }
\psfrag{i}[l][c][0.9]{$\mathbf{E}^I(t)$}
\psfrag{t}[c][c][0.9]{$\frac{d}{dt}\mathbf{E}^I(t)$}
\psfrag{s}[c][c][0.9]{$\frac{d^2}{dt^2}\mathbf{E}^I(t)$}
\includegraphics[width=0.9\columnwidth]{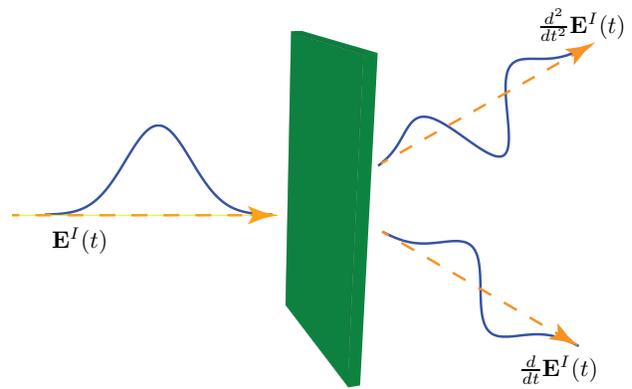}
\caption{Differentiator spacetime metasurface. A normally incident Gaussian pulse is time-differentiated and different time derivatives are transmitted to different specified directions.}
\label{fig:metasurf_diff}
\end{figure}

The second application is depicted in Fig.~\ref{fig:metasurf_diff}. It represents a metasurface that is a temporal differentiator and a spatial splitter. Specifically, the metasurface transforms a normally incident electromagnetic wave in the following fashion: it generates the first time-derivative of part of the wave and radiates the resulting wave in a specified direction, and generates the second time-derivative of the rest of the wave, that it radiates in another specified direction. In this way, detectors placed in the corresponding directions receive the two derivatives, in real-time. Such a multiple-transformation system is possible, again following the procedure described in the previous section, if the susceptibility tensors include a sufficient number of degrees of freedom~\cite{achouri2014general}.

\section{Conclusions} \label{sec:concl}

Spacetime processing metasurfaces obtained by generalized sheet transition conditions (GSTCs) susceptibility synthesis have been presented. It has been shown that such metasurfaces could perform multiple simultaneous spatio-temporal processing transformations on incident electromagnetic waves. Two examples have been shown: a time-reversal space-refractor metasurface and time-space differentiator metasurface. Such spacetime metasurfaces require active elements and may thereby also provide gain and reconfigurability~\cite{Gupta_AXV1511_11_2015}.

\vspace{5mm}

\bibliographystyle{jabbrv_ieeetr}
\bibliography{ReferenceList2}

\begin{thebibliography}{10}

\bibitem{holloway2012overview}
C.~Holloway, E.~F. Kuester, J.~Gordon, J.~O'Hara, J.~Booth, and D.~Smith, ``An
  overview of the theory and applications of metasurfaces: the two-dimensional
  equivalents of metamaterials,'' {\em\JournalTitle{IEEE Antennas Propag.
  Mag.}}, vol.~54, pp.~10--35, April 2012.

\bibitem{niemi2013synthesis}
T.~Niemi, A.~O. Karilainen, S.~Tretyakov, {\em et~al.}, ``Synthesis of
  polarization transformers,'' {\em\JournalTitle{Antennas and Propag., IEEE
  Trans. on}}, vol.~61, no.~6, pp.~3102--3111, 2013.

\bibitem{kodera2011artificial}
T.~Kodera, D.~L. Sounas, and C.~Caloz, ``Artificial faraday rotation using a
  ring metamaterial structure without static magnetic field,''
  {\em\JournalTitle{Applied Physics Letters}}, vol.~99, no.~3, p.~031114, 2011.

\bibitem{ra2013total}
Y.~Ra'di, V.~S. Asadchy, S.~Tretyakov, {\em et~al.}, ``Total absorption of
  electromagnetic waves in ultimately thin layers,'' {\em\JournalTitle{Antennas
  and Propag., IEEE Trans. on}}, vol.~61, no.~9, pp.~4606--4614, 2013.

\bibitem{salem2014manipulating}
M.~A. Salem and C.~Caloz, ``Manipulating light at distance by a metasurface
  using momentum transformation,'' {\em\JournalTitle{Optics express}}, vol.~22,
  no.~12, pp.~14530--14543, 2014.

\bibitem{achouri2015metasurface}
K.~Achouri, G.~Lavigne, M.~A. Salem, and C.~Caloz, ``Metasurface spatial
  processor for electromagnetic remote control,'' {\em\JournalTitle{arXiv
  preprint arXiv:1510.05726}}, 2015.

\bibitem{Sounas_TAP_01_2013}
D.~L. Sounas, T.~Kodera, and C.~Caloz, ``Electromagnetic modeling of a
  magnet-less non-reciprocal gyrotropic metasurface,'' {\em\JournalTitle{Trans.
  Antennas Propag.}}, vol.~61, pp.~221--231, Jan. 2013.

\bibitem{hadad2015space}
Y.~Hadad, D.~Sounas, and A.~Alu, ``Space-time gradient metasurfaces,''
  {\em\JournalTitle{Physical Review B}}, vol.~92, no.~10, p.~100304, 2015.

\bibitem{Khan_ISAP_11_2015}
B.~A. Khan, S.~Gupta, and C.~Caloz, ``Spatial nonreciprocal and nongyrotropic
  structure,'' (Hobart, Australia), pp.~154--157, Nov. 2015.

\bibitem{idemen1987boundary}
M.~Idemen and A.~H. Serbest, ``Boundary conditions of the electromagnetic
  field,'' {\em\JournalTitle{Electronics Letters}}, vol.~23, no.~13,
  pp.~704--705, 1987.

\bibitem{achouri2014general}
K.~Achouri, M.~Salem, and C.~Caloz, ``General metasurface synthesis based on
  susceptibility tensors,'' {\em\JournalTitle{IEEE Trans. Antennas Propag.}},
  2014.

\bibitem{Gupta_AXV1511_11_2015}
S.~Gupta and C.~Caloz, ``Perfect dispersive medium,''
  {\em\JournalTitle{arXiv:1511.00671}}, Nov. 2015.

\end{thebibliography}

\end{document}